# How do influential and non-influential papers spread online?

**Renmeng Cao**

WISE Lab, Faculty of Humanities and Social Sciences, Dalian University of Technology, Dalian 116085, China

## Abstract

Social media has become an important channel for publicizing academic research. Employing a dataset of about 10 million tweets of 584,264 scientific papers from 2012 to 2018, this study investigates the differential diffusion of influential and non-influential papers (divided by *Average journal impact factor percentile*). We find that non-influential papers shows a diffusion trend with multiple rounds, sparse, short-duration and small-scale bursts. In contrast, the bursts of influential journals are characterized by a small number of persistent, dense and large-scale bursts. Influential papers are generally disseminated to many loosely connected communities, while non-influential papers are diffused to several densely connected communities.

## 1. Introduction

Recently, an increasing number of journal publishers and authors are using social media to promote their research outputs, which not only boosts the social and scientific impact of scientific publications but also enhances the interactions between the scientific community and the public (Bik & Goldstein, 2013; Darling et al., 2013; Zheng et al., 2019). According to a survey by Pew research center[1], millions of users read science-related information on Facebook feeds or elsewhere on social media, which makes it a reality for scientific papers to obtain large-scale attention. Just like breaking news, scientific papers can also cumulate much attention in a very short time, which in turn creates bursty diffusion with profound effect (Cao et al., 2021; Cui et al., 2019; Zakhlebin & Horvát, 2020). As scientific research increasingly shape public discourse and impact the decision-making of both individuals and policymakers, there is a growing interest in studying how scientific findings are diffused online (Yin et al., 2021; Zakhlebin & Horvát, 2020).

### 1.3   Research questions

By constructing diffusion (retweet) networks of scientific papers, this study aims to compare the online diffusion of papers published in influential and non-influential journals. Specifically, we seek to address the following questions:

RQ1. What communication effect do papers published in different levels of journals

---

[1] https://www.pewresearch.org/science/2018/03/21/the-science-people-see-on-social-media/ Retrieved from 2021/12/07

exhibit?

RQ2. What kind of bursty trends do they exhibit?

RQ3. What are the differences in the diffusion mechanisms behind them?

Answers to these questions not only further enhance the comprehension of the diffusion process of scientific papers, but also help to improve the diffusion of scientific knowledge.

## 2 Methods

### 2.1 Defining influential and non-influential papers

Considering that the papers in our dataset are from different subject areas, the metric of the journal impact factors (IF) is not suitable to rank the journals. Alternatively, we use the *Average Journal Impact Factor (JIF) Percentile* metric[2], which is computed as follows:

$$Average\ JIF\ Percentile = \frac{JIF\ Percentile_1 + \cdots + JIF\ percentile_N}{N},$$

where $JIF\ percentile_N$ denotes the JIF Percentile of a subject area a journal belongs to, and $N$ denotes the number of subject areas a journal belongs to. Compared with impact factor, *Average JIF Percentile* improves the relative value of impact factor, has a smaller coefficient of variation and the data distribution is closer to a normal distribution, which is good for horizontal comparison among journals (Yu & Yu, 2016). Given the fluctuations of *Average JIF Percentile* at different years for the same journal, this study uses the mean of *Average JIF Percentile* between 2012 and 2018 ("$P_7$" hereafter) as a proxy for the ranking of journals. Figure 1 shows the distribution of $P_7$ values among the 300 journals. From the figure, it can be seen that these scientific journals come from different impact levels. Moreover, the distribution is very uneven, ranging from 11.08 to 99.72. The median and mean of the distribution are 76.87 and 80.02, respectively.

---

[2] http://help.incites.clarivate.com/incitesLiveJCR/glossaryAZgroup/g4/9995-TRS.html (Accessed October 12, 2021).

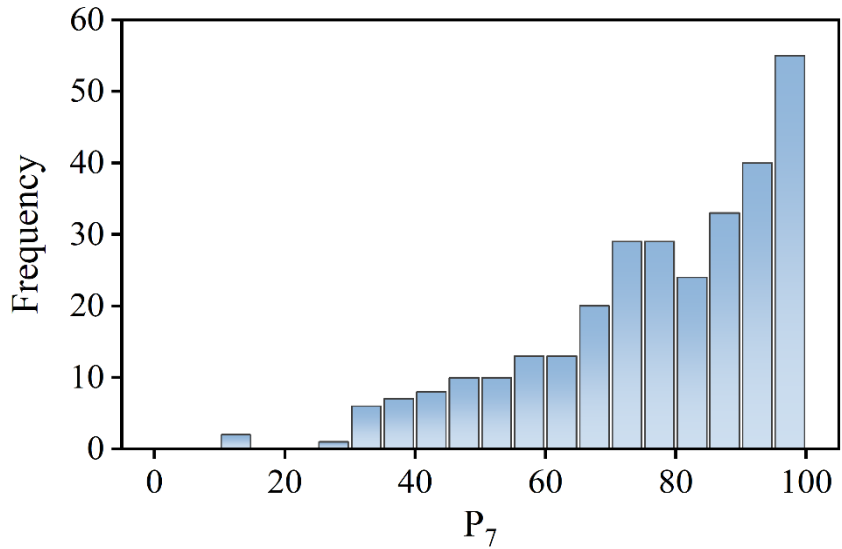

Figure 1. The distribution of $P_7$ values of the scientific journals in the dataset.

To compare the online diffusion of papers with different levels of journals, this paper pays particular attention to *highly retweeted papers* published between 2012 and 2018 on Twitter. First, we give a uniform standard of *highly retweeted papers* from different journal levels, which are defined as the papers that were retweeted larger than 100 times. Next, we rank the scientific journals of these highly retweeted papers by $P_7$ values in descending order. "Famous" journals are defined as the top 10% journals, and "non-famous" journals are defined as the journals with $P_7$ values lower than the median (80.23). Then, 30 "famous" journals and 150 "non-famous" journals are obtained. The $P_7$ values of "famous" journals range from 98.74 to 99.71, and that of "non-famous" journals are distributed between 11.80 and 80.09. The median of the former is 98.74, while that of the latter is 65.68. Finally, using the 180 journals to match scientific papers, we obtain 4,126 Influential papers and 1,158 non-influential papers. To construct a control group for the study, an equal number of influential and non-influential papers (1,158) are chosen.

## 2.2 Measuring the diffusion network

**(1) Diffusion metrics**

We measure the diffusion patterns of influential and non-influential papers from the dimensions of *Scale*, *Breadth* and *Average depth*. To exemplify the definition of these three dimensions, Figure 3 shows a sketch of a diffusion network in a tree layout. The pink node represents the root node (i.e., a scientific paper). The blue nodes are the communicators of the paper on Twitter, including both initiators and retweeters. These nodes are classified into different levels based on their shortest distance from the root node. Therefore, for a specific paper:

- *Scale* is defined as the total number of communicators of the paper;
- *Breadth* is defined as the maximum number of communicators amongst levels;
- *Average depth* is defined as the average shortest distance between the root node

and other nodes.

Take the diffusion network in Figure 3 as an example, the scale, breadth and average depth of the network are 13, 5, and 2.08, respectively.

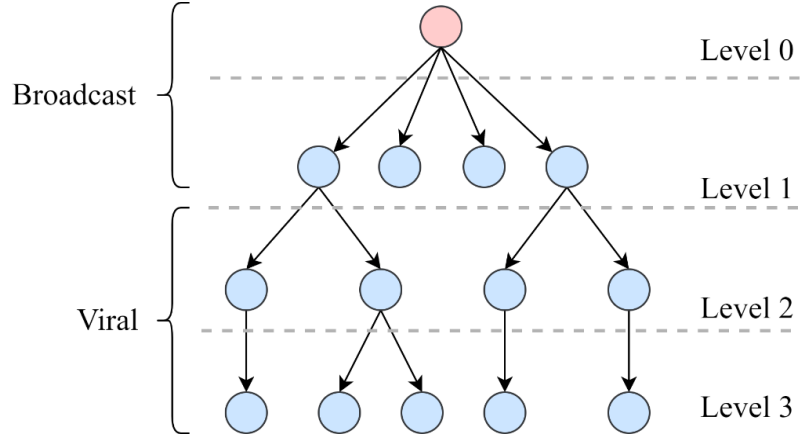

Figure 3. A schematic representation of a diffusion network in a tree layout.

**(2) Modularity**

*Modularity* is a measure of the structure of networks that evaluates the quality of network division (Newman, 2006; Newman & Girvan, 2004). It is defined as the number of edges falling within groups minus the expected number in an equivalent network with edges placed at random. The modularity is computed as follows:

$$Q = \sum_i (e_{ii} - a_i^2) \qquad (1)$$

where $e_{ii}$ represents the fraction of edges in the network that connect nodes in the same community (i.e., intra-community links), and $a_i$ represents the fraction of edges that are connected to vertices in community *i*. The range of modularity ranges from -0.5 to 1. Modularity is positive if the number of edges within a community exceeds the expected fraction in a random network. Networks with high modularity have dense connections between the nodes within modules (also called groups, clusters or communities) but sparse connections between nodes in different modules.

# 3  Results

This section consists of three parts: the first part presents the communication effect of influential and non-influential journals papers. The second part shows the bursty trends of the two categories of papers. The last one focuses on the differences in mechanisms by which influential and non-influential papers go viral.

## 3.1  Communication effect

To make a comparison of the diffusion characteristics of influential and non-influential papers, we use the complementary cumulative distribution function (CCDF) to display the distribution of variables, which describes the probability that a variable value is larger than $x$. Figure 5a plots the CCDFs of diffusion scale for influential and non-influential papers. Of Influential papers, there are 96.7% of papers less than 1000 in scale and 3.3% larger than 1000 in scale. Of non-influential papers, the proportion of papers with a scale smaller than 1000 is 96.2% and that of the papers whose diffusion scale exceeds 1000 is 3.8%. The median diffusion scale for Influential papers is 186 and that of non-influential papers is 157. In terms of breadth, the proportion of influential and non-influential papers with a breadth less than 1000 is 96.7% and 97.6%, respectively (Figure 5b). The median breadth of Influential papers is 158 and that of non-influential papers is 137. In the terms of average depth, the proportion of influential and non-influential papers with an average depth over 1.9 is 72.1% and 47.1%, respectively (Figure 5c). Figure 5d and Figure 5e show the mean average depth of influential and non-influential papers at every scale and depth, respectively. The two figures display almost the same trends: the values of mean average depth of non-influential papers are mainly distributed between 0.4 and 0.6, while that of Influential papers range from 0.1 to 0.3.

In summary, the gap of diffusion scale and breadth for influential and non-influential papers is narrow, while that of average depth for the two kinds of papers is significant. In other words, reaching the same level of scale and breadth with Influential papers need to cross a longer network distance.

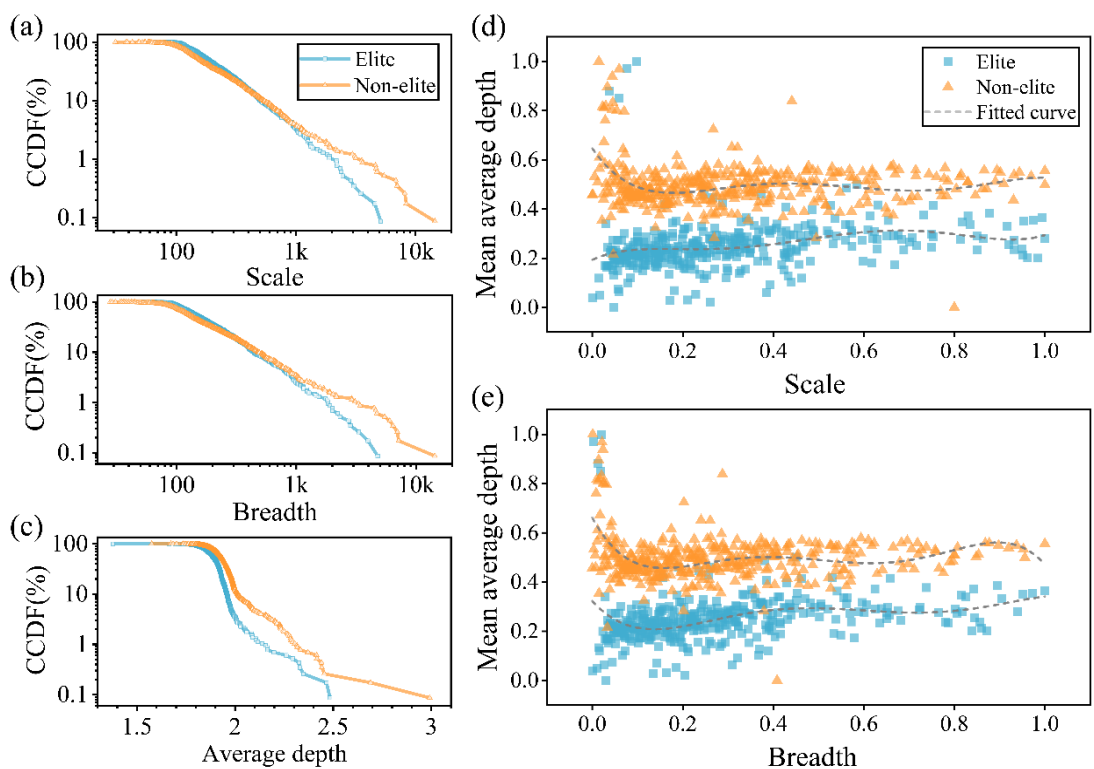

Figure 5. Complementary cumulative distribution functions (CCDFs) of diffusion network for influential and non-influential papers.

Figure 7. CCDFs of the burstiness of influential and non-influential papers: (a) The number of peaks in the diffusion curve; (b) The width of peaks; (c) The audience size of peaks; (d) The interval between two adjacent peaks

## 3.2 Diffusion mechanisms

The results in the above two subsections indicate that there are clear differences in diffusion network structure and bursty trends between influential and non-influential papers. In the following subsections, we try to disclose the underlying mechanism behind the diffusion of influential and non-influential papers and the differences in diffusion mechanisms.

**(1) Cross-community patterns**

Previous studies show that community structure plays an important role in information diffusion (Galstyan & Cohen, 2007; Gleeson, 2008; Grabowicz et al., 2012). For example, Weng et al. (2013) found that the more communities a meme permeated, the more viral it is. To compare the cross-community patterns of influential and non-influential journals, we perform community detection on diffusion networks of influential and non-influential papers by Lovin algorithms (Blondel et al., 2008) and use the metrics of *modularity* to measure the quality of the community division (Newman, 2006; Newman & Girvan, 2004). A network with high modularity has dense connections between the nodes within communities but sparse connections between nodes in different communities. Figure 8 shows the patterns of cross-community diffusion of influential and non-influential papers. Figure 8a compares the number of communities of influential and non-influential papers at each burst. From the figure, we can see that Influential papers are diffused to many communities at the first bursts, but only to a few communities in subsequent bursts. Non-influential papers are disseminated to a few communities in each burst, but they experience more rounds of bursts than the famous. Their final cumulative number of communities exceeds that of Influential papers in the subsequent bursts. Figure 8b shows the community size of influential and non-influential papers at each burst. Influential and non-influential papers are adopted by a larger number of people at the first bursts, while they receive a little attention in the subsequent bursts. The final cumulative community size of non-influential papers exceeds that of Influential papers as the number of bursts increases. We further compare the modularity and the inter-community links of influential and non-influential papers. As shown in Figure 8c, a greater fraction of non-influential papers has modularity distributing between 0 and 0.6, while a greater fraction of Influential papers has modularity ranging from 0.6 to 1.0. That is to say, Influential papers show a clearer community structure than the non-famous. Figure 8d shows the CCDFs of the number of inter-community links for influential and non-influential papers. About 31% of Influential papers have inter-community links larger than 10, while nearly 46% of non-influential papers do.

To sum up, the number and size of community for influential and non-influential papers decays as the round of bursts increases, while the final cumulative community number and size of non-influential papers are larger than that of the famous. Influential papers show a clearer community structure than the non-famous. The audience of Influential papers comes from many communities with sparse links in between, while that of non-influential papers comes from several communities with dense links in between.

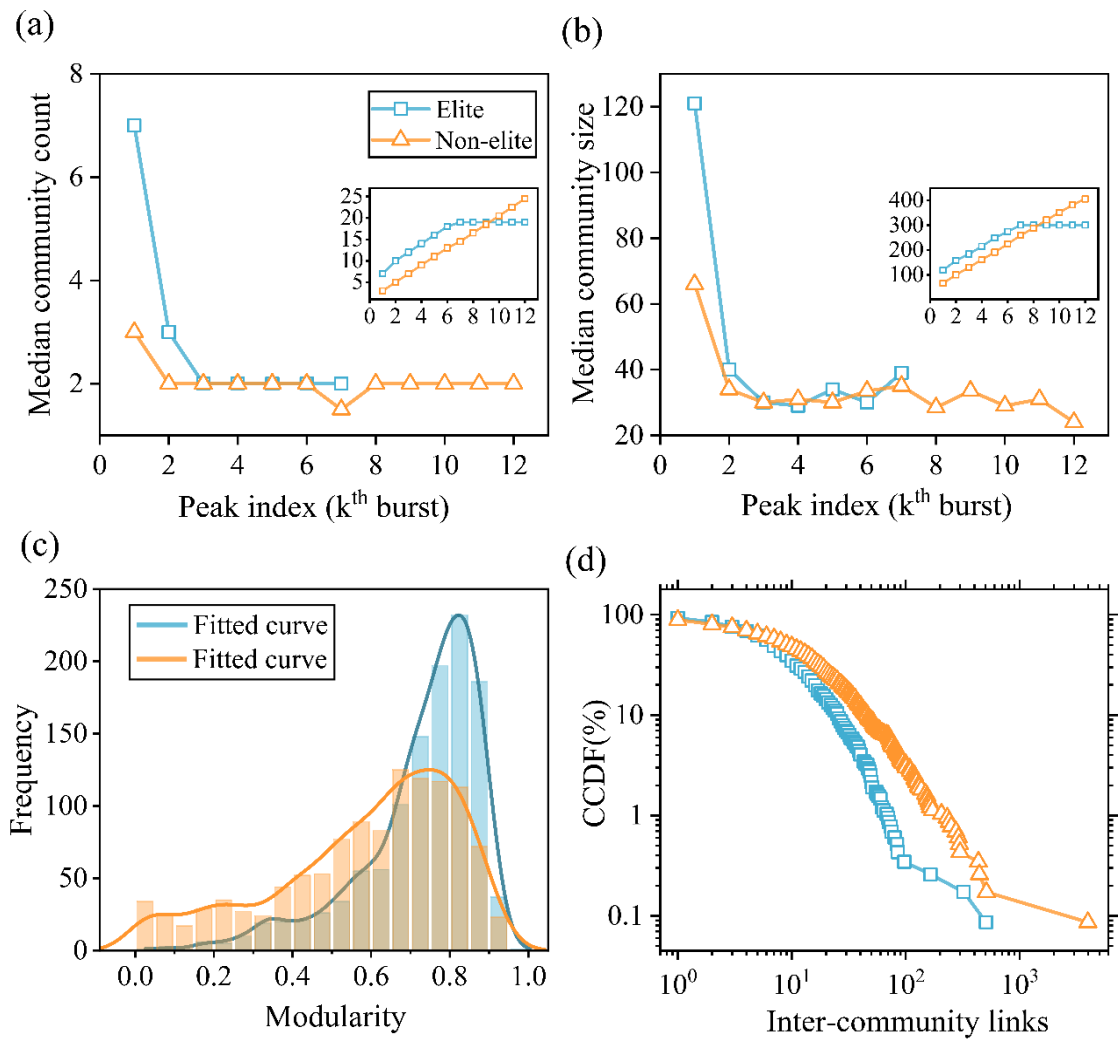

Figure 8  Patterns of cross-community of influential and non-influential papers. (a) The number of communities at the k[th] burst. (b) The community size at the kth burst. (c) The distribution of community modularity. (d) CCDFs of the number of inter-community links.

# 4  Conclusion and discussion

This study compares the online diffusion of scientific papers published in different levels of journals. The results show that in order to obtain large-scale attention, non-influential journals paper need to pay more effort than Influential papers: they need to spend longer network distance and spreading time to cumulate audience, exhibiting a diffusion trend with multiple rounds, sparse, short-duration and small-scale bursts.